\documentclass[12pt]{article}
\begin{document}

\title{\small \rm
\begin{flushright}
\small{OUTP-97-66P}\\
\end{flushright}
\vspace{2cm}
\LARGE \bf What lattice calculations tell us about the
glueball spectrum}
\date{}

\author{Michael Teper\thanks{Talk at HEP97, Jerusalem, August 1997
} 
\vspace{0.5cm} \\
{Theoretical Physics }\\
{University of Oxford  }\\
{Oxford, OX1 3NP, U.K.}}
\maketitle

\begin{abstract}
I review what lattice QCD simulations have to tell us
about the glueball spectrum. We see that the various
lattice calculations are in good agreement with each 
other. They predict that prior to mixing with
nearby flavour singlet quarkonia the lightest
glueball states are the scalar at $1.61\pm 0.15 \ {\rm GeV}$,
the tensor at $2.26\pm 0.22 \ {\rm GeV}$, and the pseudoscalar
at $2.19\pm 0.32 \ {\rm GeV}$.
\end{abstract}

\vfill
\hfill
hep-ph/9711299

\newpage

\section {Introduction} 

My topic here concerns the glueball spectrum.   
The physics question is: where, in the experimentally
determined hadron spectrum, are the glueballs hiding? 
Ideally I should be telling you what happens when you
simulate QCD with realistically light quarks. But it is
going to be a few years yet before I can do that. What
I will focus on here are lattice glueball calculations 
in the SU(3) gauge theory without quarks. We now
know what are the lightest states in the continuum
(rather than lattice) theory; and I will tell you what
they are. The next step, if we want to make contact
with the real world, is to introduce the physical
GeV mass scale. This introduces uncertainties which
I will try to estimate for you. The final step is to discuss
possible mixing scenarios with nearby flavour singlet
quarkonia. At this stage we can look at the experimental
spectrum and pinpoint 
the experimental states most likely to have large
glueball components.
I will not have time to say much
about the latter topics and refer you instead to
ref \cite{fec-mt} 
and 
ref \cite{newton-mt} 
where you will also find a more complete set of references.

The states in the pure SU(3) gauge theory are
glueballs by definition - we only have gluons in the 
theory. If you want hadrons with quarks then you can propagate
quarks in this gluonic vacuum and tie such propagators
together so that the object propagating has the appropriate
hadronic quantum numbers. That is to say, you calculate
hadron masses in the relativistic valence quark approximation
(the `quenched approximation') to QCD.
The spectrum one obtains this way is a remarkably good
approximation to the observed light
hadron spectrum. This is not too surprising: one
reason we were able to learn of the existence of quarks in the 
first place is because the low-lying hadrons are in fact well 
described by a simple valence quark picture. 

Of course in this theory, with no vacuum quark loops, we don't
have mixing between quarkonia and glueballs.
There is however reason to believe that this mixing is weak 
in the real world --
the Zweig (OZI) rule: hadron decays where the initial quarks all
have to annihilate are strongly suppressed, e.g.in $\phi$ decays. 
Such a decay may be thought
of as $quarks \to glue \to quarks$. Glueball mixing with quarks
should therefore be $\surd$OZI suppressed. As should 
glueball decays into hadrons composed of quarks. The
existence of such a suppression is supported by a recent 
lattice calculation
ref \cite{GF11-decay}.

The picture we have in mind is therefore as follows.
The glueballs will only be mildly affected by the 
presence of light quarks. They will, of course, decay into pions
etc. but their decay width will be relatively small; and there
will be a correspondingly small mass shift. Only if there
happens to be a flavour singlet quarkonium state close by
in mass will things be very different, because of the mixing
of these nearly degenerate states. In this context we
expect `close by' to mean within $\sim 100$ MeV.
So we view the glueballs in the pure SU(3) gauge theory
as being the `bare' glueballs of QCD which may mix with
nearby quarkonia to produce the hadrons that are observed
in experiments. All this is an assumption of course, albeit
well motivated. If true it tells us that the glueballs,
whether mixed with quarkonia or not, should lie close to
the masses they have in the gauge theory. So we now turn to 
the calculation of those masses.

\section {Glueballs in the SU(3) gauge theory}

In lattice calculations (Euclidean) space-time is discretised
to a hypercubic lattice, and the volume is made finite
and (anti)periodic. So the first step
is to be able to calculate masses reliably on such a
lattice hypertorus. The second is to make sure the volume
is large enough. Assuming this has been done we obtain 
the mass spectrum, $am_i(a)$, of the discretised
theory in units of the lattice spacing $a$.
What we actually want is the corresponding
spectrum of the continuum theory, $a=0$. To obtain this
we proceed as follows. Theoretically we know that
in this theory the leading lattice spacing corrections 
to dimensionless ratios of physical masses are $O(a^2)$
where $a$ is the lattice spacing
ref \cite{sym}. 
So  for small enough $a$ we can extrapolate 
our calculated mass ratios to $a=0$ using
\begin{equation}
{{a m_1(a)} \over {a m_2(a)}} \equiv
{{m_1(a)} \over {m_2(a)}} =
{{m_1(a=0)} \over {m_2(a=0)}} + c (a m)^2 
\label{A1}
\end{equation}
where $m$ may be chosen to be $m_1$ or $m_2$ or some
other physical mass: the difference between these choices
is clearly higher order in $a^2$ - which we are neglecting. 
In practice I shall use $am_2 = am = a\surd\sigma$,
where $a^2\sigma$ is the confining string tension as
calculated in lattice units, and $am_1 = am_G$ will
be a glueball mass.

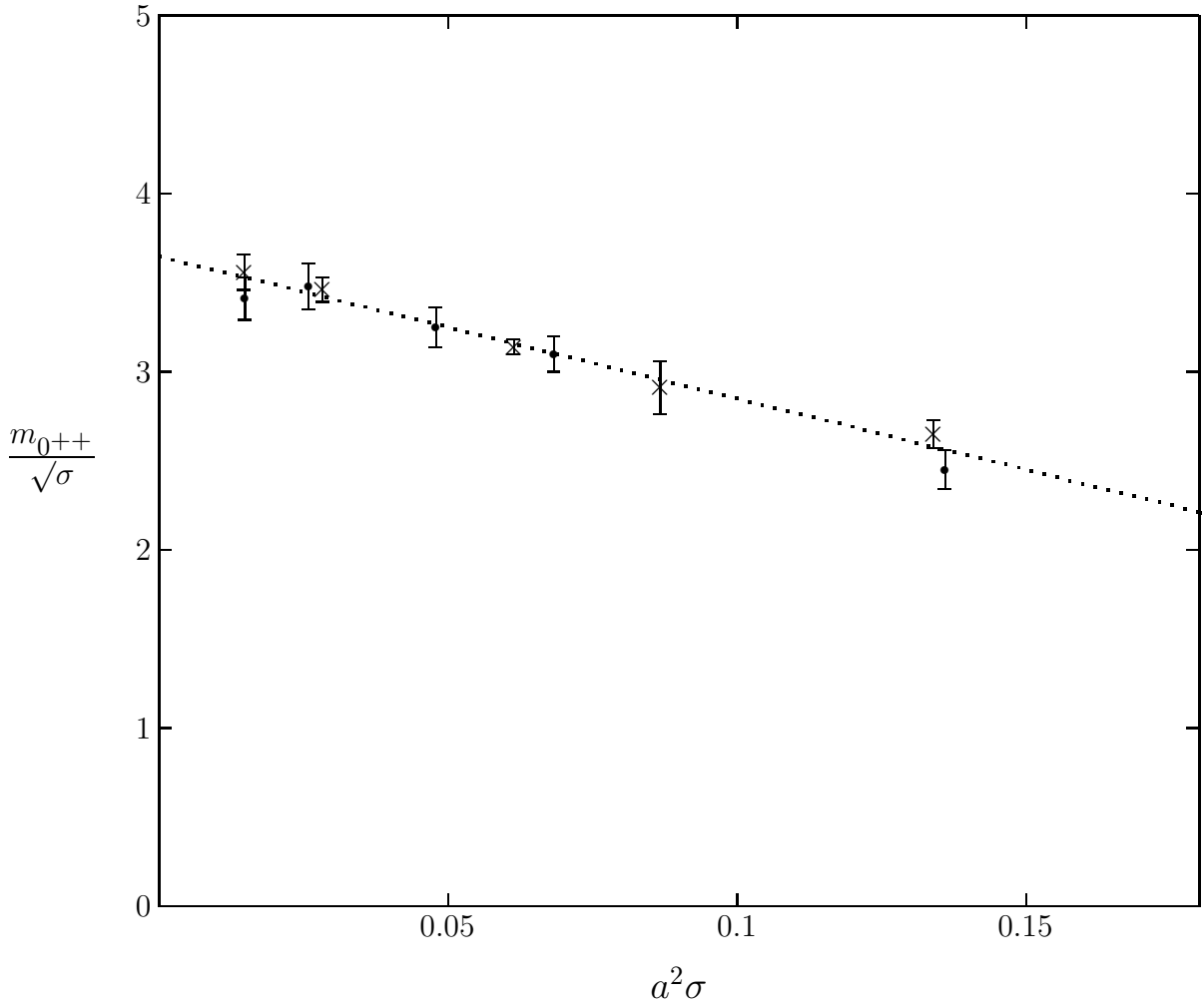
\begin{figure}[t]
\begin{flushright}
\leavevmode
\setlength{\unitlength}{0.240900pt}
\ifx\plotpoint\undefined\newsavebox{\plotpoint}\fi
\sbox{\plotpoint}{\rule[-0.200pt]{0.400pt}{0.400pt}}%
\begin{picture}(1800,1440)(0,0)
\font\gnuplot=cmr10 at 12pt
\gnuplot
\sbox{\plotpoint}{\rule[-0.200pt]{0.400pt}{0.400pt}}%
\put(120.0,31.0){\rule[-0.200pt]{4.818pt}{0.400pt}}
\put(108,31){\makebox(0,0)[r]{{$0$}}}
\put(1736.0,31.0){\rule[-0.200pt]{4.818pt}{0.400pt}}
\put(120.0,311.0){\rule[-0.200pt]{4.818pt}{0.400pt}}
\put(108,311){\makebox(0,0)[r]{{$1$}}}
\put(1736.0,311.0){\rule[-0.200pt]{4.818pt}{0.400pt}}
\put(120.0,592.0){\rule[-0.200pt]{4.818pt}{0.400pt}}
\put(108,592){\makebox(0,0)[r]{{$2$}}}
\put(1736.0,592.0){\rule[-0.200pt]{4.818pt}{0.400pt}}
\put(120.0,872.0){\rule[-0.200pt]{4.818pt}{0.400pt}}
\put(108,872){\makebox(0,0)[r]{{$3$}}}
\put(1736.0,872.0){\rule[-0.200pt]{4.818pt}{0.400pt}}
\put(120.0,1153.0){\rule[-0.200pt]{4.818pt}{0.400pt}}
\put(108,1153){\makebox(0,0)[r]{{$4$}}}
\put(1736.0,1153.0){\rule[-0.200pt]{4.818pt}{0.400pt}}
\put(120.0,1433.0){\rule[-0.200pt]{4.818pt}{0.400pt}}
\put(108,1433){\makebox(0,0)[r]{{$5$}}}
\put(1736.0,1433.0){\rule[-0.200pt]{4.818pt}{0.400pt}}
\put(574.0,31.0){\rule[-0.200pt]{0.400pt}{4.818pt}}
\put(574,19){\makebox(0,0){\shortstack{\\ \\ \\ {$0.05$}}}}
\put(574.0,1413.0){\rule[-0.200pt]{0.400pt}{4.818pt}}
\put(1029.0,31.0){\rule[-0.200pt]{0.400pt}{4.818pt}}
\put(1029,19){\makebox(0,0){\shortstack{\\ \\ \\ {$0.1$}}}}
\put(1029.0,1413.0){\rule[-0.200pt]{0.400pt}{4.818pt}}
\put(1483.0,31.0){\rule[-0.200pt]{0.400pt}{4.818pt}}
\put(1483,19){\makebox(0,0){\shortstack{\\ \\ \\ {$0.15$}}}}
\put(1483.0,1413.0){\rule[-0.200pt]{0.400pt}{4.818pt}}
\put(120.0,31.0){\rule[-0.200pt]{394.112pt}{0.400pt}}
\put(1756.0,31.0){\rule[-0.200pt]{0.400pt}{337.742pt}}
\put(120.0,1433.0){\rule[-0.200pt]{394.112pt}{0.400pt}}
\put(-48,732){\makebox(0,0){{\Large{${m_{0^{++}} \over {\surd\sigma}}$}}}}
\put(938,-89){\makebox(0,0){{\large{$a^2\sigma$}}}}
\put(120.0,31.0){\rule[-0.200pt]{0.400pt}{337.742pt}}
\put(255,987){\circle*{12}}
\put(355,1007){\circle*{12}}
\put(555,942){\circle*{12}}
\put(741,900){\circle*{12}}
\put(1356,718){\circle*{12}}
\put(255.0,954.0){\rule[-0.200pt]{0.400pt}{16.140pt}}
\put(245.0,954.0){\rule[-0.200pt]{4.818pt}{0.400pt}}
\put(245.0,1021.0){\rule[-0.200pt]{4.818pt}{0.400pt}}
\put(355.0,970.0){\rule[-0.200pt]{0.400pt}{17.586pt}}
\put(345.0,970.0){\rule[-0.200pt]{4.818pt}{0.400pt}}
\put(345.0,1043.0){\rule[-0.200pt]{4.818pt}{0.400pt}}
\put(555.0,911.0){\rule[-0.200pt]{0.400pt}{14.936pt}}
\put(545.0,911.0){\rule[-0.200pt]{4.818pt}{0.400pt}}
\put(545.0,973.0){\rule[-0.200pt]{4.818pt}{0.400pt}}
\put(741.0,872.0){\rule[-0.200pt]{0.400pt}{13.490pt}}
\put(731.0,872.0){\rule[-0.200pt]{4.818pt}{0.400pt}}
\put(731.0,928.0){\rule[-0.200pt]{4.818pt}{0.400pt}}
\put(1356.0,687.0){\rule[-0.200pt]{0.400pt}{14.936pt}}
\put(1346.0,687.0){\rule[-0.200pt]{4.818pt}{0.400pt}}
\put(1346.0,749.0){\rule[-0.200pt]{4.818pt}{0.400pt}}
\put(254,1029){\makebox(0,0){$\times$}}
\put(377,1001){\makebox(0,0){$\times$}}
\put(678,911){\makebox(0,0){$\times$}}
\put(908,847){\makebox(0,0){$\times$}}
\put(1338,774){\makebox(0,0){$\times$}}
\put(254.0,1001.0){\rule[-0.200pt]{0.400pt}{13.490pt}}
\put(244.0,1001.0){\rule[-0.200pt]{4.818pt}{0.400pt}}
\put(244.0,1057.0){\rule[-0.200pt]{4.818pt}{0.400pt}}
\put(377.0,982.0){\rule[-0.200pt]{0.400pt}{9.395pt}}
\put(367.0,982.0){\rule[-0.200pt]{4.818pt}{0.400pt}}
\put(367.0,1021.0){\rule[-0.200pt]{4.818pt}{0.400pt}}
\put(678.0,900.0){\rule[-0.200pt]{0.400pt}{5.541pt}}
\put(668.0,900.0){\rule[-0.200pt]{4.818pt}{0.400pt}}
\put(668.0,923.0){\rule[-0.200pt]{4.818pt}{0.400pt}}
\put(908.0,805.0){\rule[-0.200pt]{0.400pt}{20.236pt}}
\put(898.0,805.0){\rule[-0.200pt]{4.818pt}{0.400pt}}
\put(898.0,889.0){\rule[-0.200pt]{4.818pt}{0.400pt}}
\put(1338.0,752.0){\rule[-0.200pt]{0.400pt}{10.600pt}}
\put(1328.0,752.0){\rule[-0.200pt]{4.818pt}{0.400pt}}
\put(1328.0,796.0){\rule[-0.200pt]{4.818pt}{0.400pt}}
\sbox{\plotpoint}{\rule[-0.500pt]{1.000pt}{1.000pt}}%
\put(120,1053){\usebox{\plotpoint}}
\put(120.00,1053.00){\usebox{\plotpoint}}
\put(140.19,1048.20){\usebox{\plotpoint}}
\put(160.35,1043.27){\usebox{\plotpoint}}
\put(180.52,1038.37){\usebox{\plotpoint}}
\put(200.71,1033.54){\usebox{\plotpoint}}
\multiput(203,1033)(20.136,-5.034){0}{\usebox{\plotpoint}}
\put(220.86,1028.56){\usebox{\plotpoint}}
\put(241.04,1023.74){\usebox{\plotpoint}}
\put(261.21,1018.83){\usebox{\plotpoint}}
\put(281.37,1013.91){\usebox{\plotpoint}}
\put(301.56,1009.10){\usebox{\plotpoint}}
\multiput(302,1009)(20.136,-5.034){0}{\usebox{\plotpoint}}
\put(321.71,1004.13){\usebox{\plotpoint}}
\put(341.78,998.88){\usebox{\plotpoint}}
\put(361.80,993.46){\usebox{\plotpoint}}
\put(381.96,988.51){\usebox{\plotpoint}}
\multiput(384,988)(20.204,-4.754){0}{\usebox{\plotpoint}}
\put(402.15,983.71){\usebox{\plotpoint}}
\put(422.31,978.75){\usebox{\plotpoint}}
\put(442.51,974.00){\usebox{\plotpoint}}
\put(462.68,969.08){\usebox{\plotpoint}}
\put(482.86,964.27){\usebox{\plotpoint}}
\multiput(484,964)(20.136,-5.034){0}{\usebox{\plotpoint}}
\put(503.01,959.29){\usebox{\plotpoint}}
\put(523.20,954.45){\usebox{\plotpoint}}
\put(543.37,949.56){\usebox{\plotpoint}}
\put(563.53,944.62){\usebox{\plotpoint}}
\multiput(566,944)(19.912,-5.857){0}{\usebox{\plotpoint}}
\put(583.47,938.88){\usebox{\plotpoint}}
\put(603.62,933.91){\usebox{\plotpoint}}
\put(623.80,929.05){\usebox{\plotpoint}}
\put(643.98,924.18){\usebox{\plotpoint}}
\put(664.13,919.22){\usebox{\plotpoint}}
\multiput(665,919)(20.204,-4.754){0}{\usebox{\plotpoint}}
\put(684.32,914.42){\usebox{\plotpoint}}
\put(704.48,909.48){\usebox{\plotpoint}}
\put(724.65,904.59){\usebox{\plotpoint}}
\put(744.83,899.75){\usebox{\plotpoint}}
\multiput(748,899)(20.136,-5.034){0}{\usebox{\plotpoint}}
\put(764.98,894.77){\usebox{\plotpoint}}
\put(785.19,890.02){\usebox{\plotpoint}}
\put(805.25,884.74){\usebox{\plotpoint}}
\put(825.28,879.35){\usebox{\plotpoint}}
\put(845.43,874.39){\usebox{\plotpoint}}
\multiput(847,874)(20.204,-4.754){0}{\usebox{\plotpoint}}
\put(865.62,869.59){\usebox{\plotpoint}}
\put(885.78,864.64){\usebox{\plotpoint}}
\put(905.95,859.76){\usebox{\plotpoint}}
\put(926.13,854.91){\usebox{\plotpoint}}
\multiput(930,854)(20.136,-5.034){0}{\usebox{\plotpoint}}
\put(946.28,849.93){\usebox{\plotpoint}}
\put(966.47,845.13){\usebox{\plotpoint}}
\put(986.64,840.20){\usebox{\plotpoint}}
\put(1006.80,835.30){\usebox{\plotpoint}}
\put(1026.99,830.47){\usebox{\plotpoint}}
\multiput(1029,830)(19.811,-6.191){0}{\usebox{\plotpoint}}
\put(1046.88,824.56){\usebox{\plotpoint}}
\put(1067.06,819.73){\usebox{\plotpoint}}
\put(1087.23,814.83){\usebox{\plotpoint}}
\put(1107.43,810.07){\usebox{\plotpoint}}
\put(1127.58,805.10){\usebox{\plotpoint}}
\multiput(1128,805)(20.204,-4.754){0}{\usebox{\plotpoint}}
\put(1147.78,800.31){\usebox{\plotpoint}}
\put(1167.94,795.37){\usebox{\plotpoint}}
\put(1188.11,790.47){\usebox{\plotpoint}}
\put(1208.29,785.64){\usebox{\plotpoint}}
\multiput(1211,785)(20.136,-5.034){0}{\usebox{\plotpoint}}
\put(1228.44,780.66){\usebox{\plotpoint}}
\put(1248.63,775.84){\usebox{\plotpoint}}
\put(1268.67,770.45){\usebox{\plotpoint}}
\put(1288.71,765.07){\usebox{\plotpoint}}
\put(1308.90,760.26){\usebox{\plotpoint}}
\multiput(1310,760)(20.136,-5.034){0}{\usebox{\plotpoint}}
\put(1329.05,755.28){\usebox{\plotpoint}}
\put(1349.23,750.44){\usebox{\plotpoint}}
\put(1369.40,745.55){\usebox{\plotpoint}}
\put(1389.56,740.61){\usebox{\plotpoint}}
\multiput(1392,740)(20.204,-4.754){0}{\usebox{\plotpoint}}
\put(1409.75,735.81){\usebox{\plotpoint}}
\put(1429.90,730.85){\usebox{\plotpoint}}
\put(1450.11,726.09){\usebox{\plotpoint}}
\put(1470.27,721.18){\usebox{\plotpoint}}
\put(1490.46,716.36){\usebox{\plotpoint}}
\multiput(1492,716)(19.811,-6.191){0}{\usebox{\plotpoint}}
\put(1510.35,710.45){\usebox{\plotpoint}}
\put(1530.53,705.62){\usebox{\plotpoint}}
\put(1550.70,700.72){\usebox{\plotpoint}}
\put(1570.86,695.78){\usebox{\plotpoint}}
\multiput(1574,695)(20.204,-4.754){0}{\usebox{\plotpoint}}
\put(1591.05,690.99){\usebox{\plotpoint}}
\put(1611.20,686.01){\usebox{\plotpoint}}
\put(1631.38,681.15){\usebox{\plotpoint}}
\put(1651.56,676.28){\usebox{\plotpoint}}
\put(1671.71,671.32){\usebox{\plotpoint}}
\multiput(1673,671)(20.204,-4.754){0}{\usebox{\plotpoint}}
\put(1691.91,666.52){\usebox{\plotpoint}}
\put(1712.06,661.57){\usebox{\plotpoint}}
\put(1732.08,656.16){\usebox{\plotpoint}}
\put(1752.15,650.91){\usebox{\plotpoint}}
\put(1756,650){\usebox{\plotpoint}}
\end{picture}
\end{flushright}
\vskip 0.15in
\caption{The scalar glueball mass: the GF11 values ($\times$)
and the rest ($\bullet$).  
The best linear extrapolation to the continuum limit
is shown.}
\label{fig-glue}
\end{figure}

In 
Fig \ref{fig-glue} 
I show how the calculated mass ratios, for the 
lightest scalar glueball, vary with $a^2\sigma$.
As you see, the behaviour is linear - not surprising given
the fact that $a^2\sigma$ is indeed small for the
values plotted. One fits a straight line and obtains
the continuum limit as the intercept at $a=0$, i.e.
at $a^2\sigma =0$. We obtain in this way the following
lightest three masses in the $continuum$ limit:
\begin{equation}
{{m_{0^{++}}}\over \surd\sigma} = 3.65 \pm 0.11 
\label{A2}
\end{equation}       
\begin{equation}
{{m_{2^{++}}}\over \surd\sigma} = 5.15 \pm 0.21
\label{A3}
\end{equation}      
\begin{equation}
{{m_{0^{-+}}}\over \surd\sigma} = 4.97 \pm 0.58
\label{A4}
\end{equation}      
Although we do not have continuum extrapolations
for other masses, the UKQCD lattice results
strongly suggest that glueballs
with other $J^{PC}$ are heavier
ref \cite{ukqcd}. 

The values of the glueball masses that we have used are
from 
refs \cite{ukqcd,GF11-G1,cm-mt,deF-G} 
and for our sources of the string tension see
refs \cite{fec-mt,newton-mt}. You may recall that
a couple of years ago some publicity was given
to an apparent discrepancy between the GF11 and UKQCD
predictions for the $0^{++}$ glueball mass. However,
as you can explicitly see in Fig 1, the various
calculations are entirely in agreement with each other.
The discrepancy was an illusion: it arose largely 
from different ways of setting the MeV scale. Such
differences should be part of the final systematic
error on the mass: as below.

\section {Glueballs masses in GeV units}

To introduce MeV units we need the string tension in these
units. We can do this by calculating the mass of the
$\rho$ or nucleon or ... in the quenched approximation
and extrapolate $m_{\rho}/\surd\sigma$ or 
$m_{N}/\surd\sigma$ or ... to the continuum limit as
we did for the glueballs. We then set $m_{\rho} =770$ MeV
or $m_N = 930$ MeV or ... to obtain $a\surd\sigma$
in MeV units. Because the quenched spectrum
differs slightly from the real world, these estimates
differ slightly. This forms part of the error. One can 
estimate the error {\it ad nauseam}, as in
refs \cite{fec-mt,newton-mt}, and this leads
to an estimate 
\begin{equation}
 \surd\sigma = 440 \pm 15 \pm 35 \ {\rm MeV} 
\label{A5}
\end{equation}       
where the first error is statistical and the second
is systematic.

We can now use this value in eqns~\ref{A2}-~\ref{A4} to express our
glueball masses in GeV units. We obtain
\begin{equation}
m_{0^{++}} = 1.61 \pm 0.07 \pm 0.13 \ {\rm GeV} 
\label{A6}
\end{equation}         
\begin{equation}
m_{2^{++}} = 2.26 \pm 0.12 \pm 0.18 \ {\rm GeV} 
\label{A7}
\end{equation}        
\begin{equation}
m_{0^{-+}} = 2.19 \pm 0.26 \pm 0.18 \ {\rm GeV} 
\label{A8}
\end{equation}        

These, then, are our best lattice predictions
for the lightest glueballs prior to any mixing with
nearby quarkonium states. In the case of the $0^{++}$
the focus is naturally on the $f_0(1500)$ and any
scalar lurking in the $f_J(1700)$. The tensor
focus is naturally on the $f_2(1900)$ and the
$G(2150)$. With the pseudoscalar things are murky:
the lattice calculation has huge errors and
is far from the obvious $\iota(1490)$ candidate;
but in this case topology is important and that
is a quantity that is sensitive to light quarks in
the vacuum.

\end{document}